\documentclass[12pt,letterpaper]{article}
\usepackage{opex3}
\usepackage{cite}

\begin{document}

\newcommand{\dIdV}{${\partial^2 I_D} / {\partial V_{D}^2}$ \, }
\newcommand{\dIdT}{${\partial I_D} / {\partial T}$ \, }
\newcommand{\diQCL}{$\delta I_{QCL}$}

\title{Rectified diode response of a multimode quantum cascade laser integrated terahertz transceiver}

\author{Gregory C. Dyer, $^{1}$ Christopher D. Norquist,$^{1}$ Michael J. Cich,$^{1,2}$ \\ Albert D. Grine,$^{1}$ Charles T. Fuller,$^{1}$ John L. Reno,$^{1}$\\ and Michael C. Wanke$^{1,*}$}

\address{$^{1}$Sandia National Laboratories, P.O. Box 5800, Albuquerque, NM 87185, USA\\
		$^{2}$Now at Soraa, Fremont, CA, 94555, USA}


\begin{abstract*}
We characterized the DC transport response of a diode embedded in a THz quantum cascade laser as the laser current was changed. The overall response is described by parallel contributions from the rectification of the laser field due to the non-linearity of the diode I-V and from thermally activated transport. Sudden jumps in the diode response when the laser changes from single mode to multi-mode operation, with no corresponding jumps in output power, suggest that the coupling between the diode and laser field depends on the spatial distribution of internal fields. The results demonstrate conclusively that the internal laser field couples directly to the integrated diode.\\
\end{abstract*}





\section{Introduction}
The introduction and subsequent development of the terahertz (THz) quantum cascade laser (QCL) has provided a coherent, high power ($>$mW) source in the 1-5 THz band of the electromagnetic spectrum.\cite{Kohler:2002ux, Williams:2007eb} A variety of imaging \cite{Barbieri:2005tj, Kim:2006cc, Lee:2006fb, Behnken:2008vc} and spectroscopic \cite{Shen:2005bi, Huebers:2006et} applications have emerged where the THz QCL functions as a local oscillator (LO) for a heterodyne receiver.  Typically such approaches require free-space coupling to a heterodyne mixer such as a Schottky diode \cite{Barbieri:2004ut, Hubers:2005ui, Lee:2008fk} or bolometer.\cite{Gao:2005ke, Richter:2008fj, Baryshev:2006hm} However, embedding a diode into a QCL ridge waveguide appears to enable direct coupling between the LO's internal field and the diode.\cite{Wanke:2010ju, Wanke:2011wt} Such a monolithically integrated THz heterodyne receiver, or THz transceiver, allows for direct measurement of the difference frequency signal generated by mixing of cavity Fabry-Perot (F-P) modes, heterodyne reception of and mixing with an external THz source,\cite{Wanke:2010ju}  locking of the F-P mode difference frequency via a phase-locked loop, characterization of the absolute frequency stability,\cite{Wanke:2011wt}  observation of feedback from an external cavity, and imaging via a feedback mechanism.\cite{Wanke:2011to} 

Previous measurements on integrated THz transceivers focused on the difference frequency generated by mixing of two or more distinct THz modes, but exactly how the laser coupled to the diode to produce the beat frequency was not fully understood.  The difference frequency between internal F-P QCL modes has also been observed in QCLs without an integrated diode.\cite{Gellie:10} In this latter diode-free case, the signal was attributed to rectification by the non-linear current-voltage characteristic (I-V) of the laser, raising the question of whether the signal measured across the terminals of an embedded diode was generated by mixing of THz fields coupled to the diode or coupling of the beat signal produced by the laser to the diode.  Determining whether the THz fields couple to the diode directly will clarify future avenues for optimizing THz transceivers. 

In this letter we characterize the direct current (DC) change to the integrated diode I-V as a function of laser current. We attribute the change to rectification of the THz fields coupled to the diode (rectified response) and to thermally activated current due to temperature changes of the laser (thermal response). In contrast to the difference frequency signal which appears only when two or more modes are present, the rectified response appears immediately at laser threshold.  The magnitude of the rectified response is proportional to the second derivative of the diode current-voltage (I-V) characteristic rather than the laser I-V, providing unequivocal evidence that the QCL laser field couples directly to the diode. 
%
%
 Sudden jumps observed in the DC diode response when the laser changes from single mode to multi-mode operation also suggest that the spatial distribution of the internal fields and the relative position of the diode affect the laser-diode coupling.
 
 \begin{figure}[htbp]
 \centering
 \includegraphics[keepaspectratio=true,width=5.2 in]{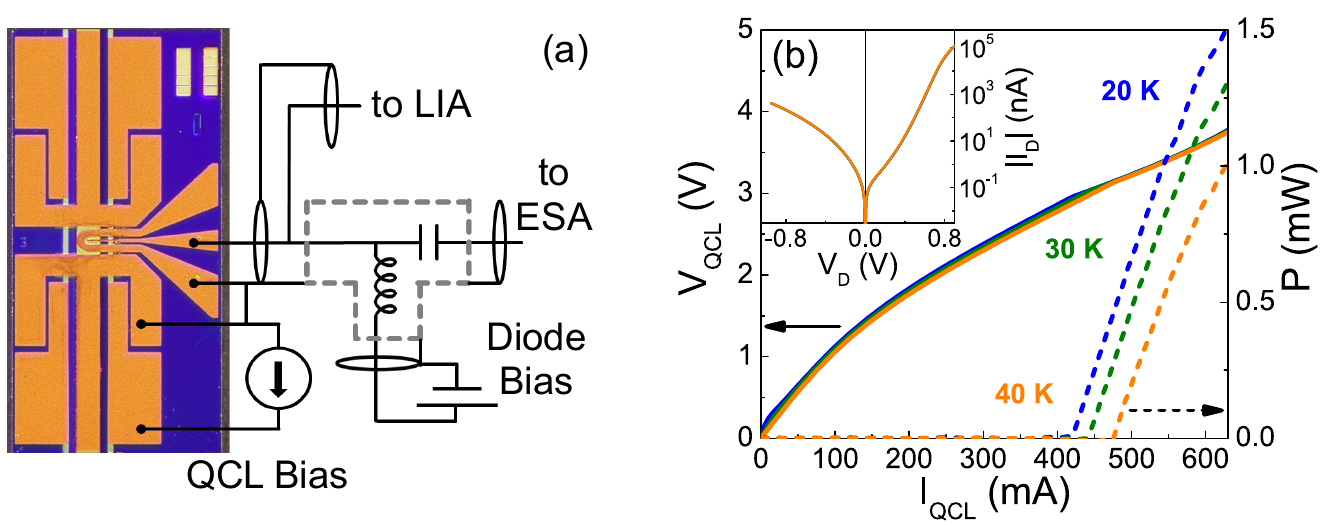}
\caption{(a) Micrograph of the 170 $\mu$m wide by 3 mm long transceiver with diode embedded in QCL ridge waveguide. The measurement circuit for diode rectification and IF signals is shown with the dashed grey region indicating a bias tee.  (b) LIV at 20 K (blue), 30 K (green) and 40 K (orange).  The diode I-V curves at all three temperatures are shown in the inset, but are virtually identical.}\label{fig:Fig1}
\end{figure}

\section{Sample and experiment description}
The THz QCL transceiver \cite{Wanke:2010ju} shown in Fig.\ \ref{fig:Fig1}(a) is based on a GaAs/Al$_{x}$Ga$_{1-x}$As QCL heterostructure (Sandia wafer VB0166).  This 2.8 THz QCL is capable of generating greater than 1 W of optical power at the operating temperatures studied as shown in Fig.\ \ref{fig:Fig1}(b).  The 1 $\mu$m diameter diode is placed near the center of the 170 $\mu$m wide by 3 mm long laser waveguide in a small opening in the top laser contact.  In order to couple the RF response of the diode from the integrated transceiver to the external measurement circuit, the diode is connected to a coplanar waveguide. The outer conductors of this RF waveguide are tied to the top contact for the laser and are referenced to ground through the RF chain. The laser power supply is floated to ensure the laser current returns on the laser bias lines and does not influence the diode bias, effectively decoupling fluctuations of the DC laser bias from the diode bias.

The circuit used to characterize the integrated diode is diagrammed in Fig.\ \ref{fig:Fig1}(a).  The QCL is biased by an ILX Lightwave LDX-3232 current supply, while the diode is  voltage biased by a Keithley 238 source-measure unit.  DC diode bias is applied through the low pass arm of a bias tee.  The IF signal generated by the diode when the laser has two or more modes passes through the high pass arm of the bias tee to a chain of RF amplifiers with a minimum of 50 dB of total gain over the frequency range of 1-18 GHz and is measured using a HP 8565E spectrum analyzer. Only the mixing term near 12.8 GHz corresponding to mixing of nearest-neighbor F-P modes was monitored. The change in the diode DC I-V  as a function of laser current was measured in two ways: (a) by comparing the diode current between different static laser currents and (b) by measuring the amplitude of the diode voltage modulation in response to a sinusoidally modulated laser current. In the former, the Stanford Research (SR) 830 lock-in amplifier shown in the circuit schematic was not connected.  In the latter, the sine wave output of the SR 830's internal oscillator was connected to the LDX-3232's modulation port, producing a modulation in the laser current about the DC set point with a measured transfer function of 162.3 mA/V.  The SR 830 signal input port was connected to the diode via a standard SMA coaxial tee in front of the RF bias tee to enable measurement of the AC component of the diode voltage at the modulation frequency.  The optical power in Fig.\ \ref{fig:Fig1}(b) was measured using an Ophir Vega Laser Power Meter and calibrated absorbing power head (PE-1).  Except when measuring the laser output power or laser emission spectrum, the emitted THz QCL power outside the cryostat was dumped. This mitigated perturbation of the laser emission spectrum due to back-reflections feeding back into the laser.\cite{Wanke:2011to}

 \begin{figure}[hbtp]
  \centering
 \includegraphics[keepaspectratio=true,width=5.2 in]{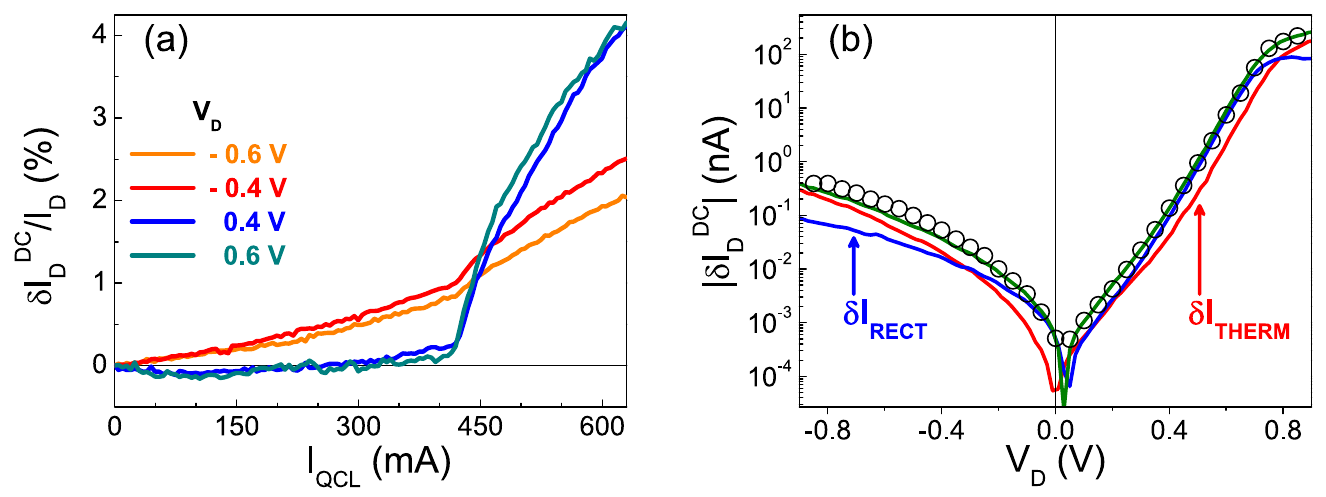}
\caption{(a) The percent change in DC diode current as a function of QCL bias current at 20 K for diode bias voltages of -0.6 V (orange), -0.4 V (red), +0.4 V (blue) and +0.6 V (teal).  (b) The DC diode response (black circles) at 20 K with 450 mA laser current is compared to the second derivative of the diode I-V (blue line), the temperature dependence of the diode I-V (red line), and a fit (green line) to the second derivative and temperature dependence of the diode I-V as a function of diode voltage bias.  These three curves represent terms in Eq.(\ref{Eq:TSE}) that are described in the text.}
\label{fig:Fig2}
\end{figure}

\section{Diode response to quasi-static laser currents}
For the first measurement method, where we compare the diode currents observed between different static laser currents, we define the diode response as 
\begin{equation}
\delta  I_{D}^{\, DC}\left( V_{D}, I_{QCL} \right) = I_{D} \left( V_{D}, I_{QCL} \right) - I_{D} ( V_{D}, I_{QCL}^{\, 0} ),
\label{Eq:QSDR} 
\end{equation}
 where $I_{QCL}$ is the laser current and $V_D$ is the diode bias. Here $I_D$ is the diode current and $I_{QCL}^{\,0}$ is an arbitrarily chosen reference laser current. In this paper we set $I_{QCL}^{\, 0} = 0$ mA for the measurement plotted in Fig.\ \ref{fig:Fig2}(a), and $I_{QCL}^{\, 0} = 425$ mA for the measurements shown in Fig.\ \ref{fig:Fig2}(b). Since the diode current changes by orders of magnitude over the range of diode biases, it is convienent to compare the fractional diode response, $\delta  I_{D}^{\, DC}\left( V_{D}, I_{QCL} \right) / I_{D} ( V_{D}, I_{QCL}^{\, 0} )$. The fractional response for 4 different diode biases as a function of the laser current is shown in Fig.\ \ref{fig:Fig2}(a).  For the positive diode biases $V_D= +0.40$ V and $V_D= +0.60$ V, the fractional diode response as a function of laser current is negligible until a sharp increase occurs at lasing threshold ($I_{QCL}=425$ mA), indicating a direct correlation between the diode response and the power of the internal QCL field.  A sharp increase of the fractional diode response at threshold is also observed under the negative diode biases $V_D= -0.40$ V and $V_D= -0.60$ V, but is confounded by an additional contribution to diode response. This additional response increases monotonically with QCL current even before the laser turns on, indicating that it is caused by something other than the laser field.

To explain the origin of both contributions, it is instructive to consider the basic current expression for an ideal diode,  $I_{D} \left( V,T \right)  = I_0 \left( \exp\left[ qV/kT \right] - 1 \right)$, where $q$ is the electron charge, $V$ is the voltage applied to the diode, $k$ is Boltzman's constant and $T$ is the temperature. As can be seen in this expression, (and in more accurate expressions\cite{Hudait:2001wa}), changes in either the voltage or temperature will change the diode current. The DC laser bias is decoupled from the diode bias, so changing the laser current should not change the DC diode bias. However, we assume that when the laser is on, the laser fields induces an alternating voltage on the diode at the laser frequency. This results in a time dependent change in diode voltage,  $\Delta V(t) = \delta V_{THz} \cos(\omega_{THz} t)$, where $\delta V_{THz}$ is the amplitude of the induced voltage. 
Similarly, even though the heat sink temperature is held constant for these measurements,  the diode temperature will change as the laser current changes. Since the diode is placed on the laser and not directly on the heat sink, the diode temperature increases as the ohmic power dissipated in the laser increases with increasing laser current. The change in temperature between two laser biases can be expressed as  $\Delta T = \delta P_{QCL} \Theta_T$ where $\delta P_{QCL}$ is the change in ohmic power dissipated in the laser and $\Theta_{T}$ is the transceiver thermal impedance that includes the heat capacitance and thermal conductance between the laser and heat sink. 

Considering both temperature and voltage variations as functions of laser current, we can re-express the change in diode current between two different laser currents as 
\begin{equation}
\delta I = I(V_D+\Delta V, T + \Delta T) - I (V_D,T). 
\label{Eq:basicTSE}
\end{equation}
%
%
Expanding the expression in a Taylor series expansion, taking the time average to obtain the DC response, and keeping only the lowest order terms, we obtain
\begin{equation}
\delta I_D^{\,DC}  =  \frac{1}{4} \frac{\partial^2 I_D}{\partial V_{D}^2}  \delta V_{THz}^2  + \frac{\partial I_D}{\partial T} \delta P_{QCL} \Theta_{T}.
\label{Eq:TSE}
\end{equation}

The first term on the right represents the change in the DC diode current due to rectification when a THz field drives the diode. Since $\delta V_{THz}^2$ is proportional to the laser power coupled to the integrated diode, this term is zero unless the laser is above threshold. The second term on the right represents the thermal contribution to the diode current and will increase as the laser current increases, independent of whether the laser field exists. 

Assuming the two terms in Eq.(\ref{Eq:TSE}) accurately represents the diode response, it appears that the thermal contribution is negligible in comparison to the rectified response for the forward biased diode curves in  Fig.\ \ref{fig:Fig2}(a).  In contrast, the thermal contribution appears to have a similar or larger magnitude compared to the rectified response for the reverse biased diode curves. Although $\Theta_T$ \cite{Vitiello:2006bs} and $\delta P_{QCL}$ can be measured, we cannot fit the data in Fig.\ \ref{fig:Fig2}(a) using Eq.(\ref{Eq:TSE}), since $\delta V_{THz}$ is an unknown function of $I_{QCL}$. 

 Although not explicitly written out in Eq.(\ref{Eq:TSE}), the partial derivatives depend on $V_D$ but are independent of $I_{QCL}$ in contrast to $\delta V_{THz}$ and $\delta P_{QCL}$ which are functions of $I_{QCL}$. Unlike $\delta V_{THz}$, the functional forms of the partial derivatives are easily obtained from the DC I-V measurement of the diode shown in the inset of Fig.\ \ref{fig:Fig1}(b).  Therefore, a  measurement of $\delta I_D$ as a function of $V_D$ with $I_{QCL}$ fixed (and hence $\delta V_{THz}$ and $\delta P_{QCL} \Theta_T$ constant), can be fit to Eq.(\ref{Eq:TSE}). This allows us to extract the values of $\delta V_{THz}$ and $\Theta_T$.  In Fig.\ \ref{fig:Fig2}(b),  $\delta I_D$ as a function of diode bias (circles) is plotted for $I_{QCL} = 450$ mA (above threshold) with $I_{QCL}^{\,0} = 425$ mA (just below threshold). Also plotted are the partial derivatives, \dIdV (blue) and \dIdT (red), calculated from the diode I-V taken at 20K, and a least-squares fit (green) of $\delta I_{D}^{\,DC}$  to Eq.(\ref{Eq:TSE}) using $\delta V_{THz}^2$ and $\Theta_T$ as fit parameters.  The remaining term, $\delta P_{QCL}$, is calculated directly from the laser I-V shown in Fig.\ \ref{fig:Fig1}(b).  The fit yielded $\delta V_{THz}^2 = 6.92 \times 10^{-5} \pm 0.35 \times 10^{-5} \, V^2$ and $\Theta_T = 12.28 \pm 0.53 \, W/K$. As seen in Fig.\ \ref{fig:Fig2}(b), the resulting overall fit (green) is in excellent agreement with the data.  In addition, the thermal impedance extracted from the fit is within a factor of 2 of the values obtained for similar lasers in the literature \cite{Vitiello:2006bs}. 

While the shapes of the red and blue curves in Fig.\ \ref{fig:Fig2}(b) reflect the functional form of the derivatives of the DC I-V of the diode, they are scaled by the fit parameters so that the blue curve represents the rectified response term in Eq.(\ref{Eq:TSE}) and the red curve represents the thermal response term. The green curve is just the sum of the two curves. Plotted this way we can immediately see that neither term by itself matches the data. In addition,  we can see which term dominates for particular values of $V_D$;   the thermally activated response is larger than the rectified response when $V_D < -0.3$ V, whereas the rectified response is larger than the thermal response when $0.3 < V_D < 0.8$ V. This result is in agreement with the behavior shown in Fig.\ \ref{fig:Fig2}(a). Since the data is well represented by Eq.(\ref{Eq:TSE}) and the rectified response term is dominant for many diode biases, we can conclude that $\delta V_{THz}$ is non-zero and therefore that the laser fields couple directly to the diode. 

 \begin{figure}[htbp]
  \centering
 \includegraphics[keepaspectratio=true,width=5.2 in]{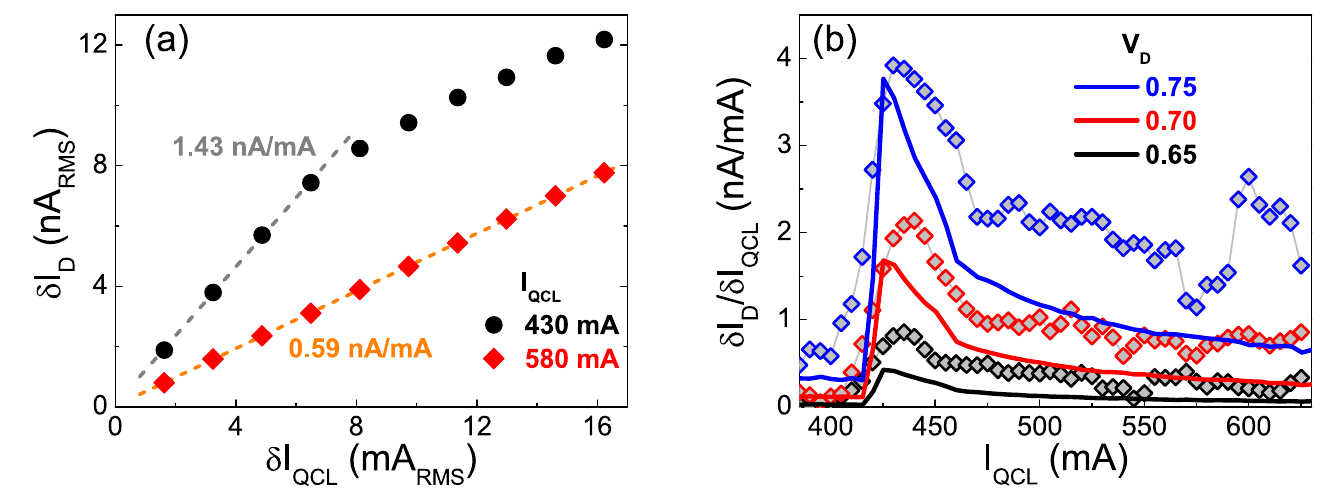}
\caption{(a) The AC rectified diode response at 20 K as a function of QCL current modulation amplitude.  The QCL current was modulated around 430 mA (black circles) and 580 mA (red diamonds).  (b) The DC and AC diode responses at 20 K are compared at three different diode bias points, 0.60 V (black), 0.65 V (red), and 0.75 V (blue).  Solid lines are the AC change in diode current.  Diamonds are the calculated derivative of the DC change in diode current.}
\label{fig:Fig3}
\end{figure}

\section{Diode response to modulated laser currents}
From Fig.\ \ref{fig:Fig1}(b) we can see that the laser power depends nearly linearly on the QCL current above threshold. If we assume that $\delta V_{THz}^2$ is proportional to the laser output power, then it too depends linearly on $\delta I_{QCL}$. In comparison, $\delta P_{QCL} ( = \delta V_{QCL}  \delta I_{QCL}) $ is proportional to $\delta I_{QCL}^2$. Therefore to minimize the relative important of the thermal contribution we want to minimize the modulation of $I_{QCL}$.  Thus  additional measurements  were performed by modulating the QCL current with small amplitude about a fixed DC bias point taking advantage of lock-in amplification to capture the small response.  Choosing  a sufficiently small QCL modulation amplitude \diQCL\ results in a time-dependent power dissipation on the chip, 
\begin{equation}
\delta P(t) = \left[ V_{QCL} + I_{QCL} \left( \partial V_{QCL}  / \partial I_{QCL} \right) \right] \delta I_{QCL} \cos \left(\omega_{mod}t \right),
\label{}
\end{equation}
that is linear in \diQCL.  In contrast, $\Delta I_{QCL}$ in the quasi-static measurements above were large resulting in quadratic and higher order terms in the power dissipation.   The root-mean-square (RMS) diode rectified current, $\delta I_D (\delta I_{QCL})$, at a QCL modulation frequency of  1.0000 kHz is shown in Fig.\ \ref{fig:Fig3}(a).  The diode bias is $V_D= +0.70$ V and the heat sink temperature is 20 K.  Instead of measuring the modulation of the diode current directly, the lock-in measured the modulation of the voltage across the diode, $\delta V_D$. From this the RMS change in diode current is calculated using $\delta I_{D} (I_{QCL}) = (\partial I_D / \partial V_D ) \delta V_D(I_{QCL}) $. The differential conductance $\partial I_D / \partial V_D$ is calculated from the static diode I-V in Fig.\ \ref{fig:Fig1}(b).  Here  we define the modulated diode response  as the instantaneous slope of the RMS change in diode current as a function of the DC QCL current bias, i.e. $\delta I_D^{AC} (I_{QCL}) / \delta I_{QCL}$. At a base laser bias current of $I_{QCL}=580$ mA (significantly above threshold), the slope is a constant 0.59 nA/mA for QCL modulation amplitudes up to at least 16 mA RMS.  For $I_{QCL}=430$ mA (just above threshold), a larger response of 1.43 nA/mA is observed for small modulation amplitudes, but the response deviates from linearity and decreases when the modulation amplitude exceeds 6 mA. We attribute the rollover to suppression of lasing when the modulation drives the laser current below threshold ($I_{QCL}=425$ mA) which reduces the peak to peak  amplitude of  $\delta I_D^{AC} (I_{QCL}) $.

The quasi-static diode response $\delta I_D^{\, DC}(I_{QCL})$ and the modulated diode response, $\delta I_D^{AC} (I_{QCL}) / \delta I_{QCL}$, can be compared by recognizing that in the limit of $\delta I_{QCL} \rightarrow 0$ the latter represents the instantaneous slope of the former.  We approximate the instantaneous slope of the DC diode response as 
\begin{equation}
\delta I_D^{AC} (I_{QCL}) / \delta I_{QCL} = \left[ I_D \left( I_{QCL} +\frac{\delta I_{QCL}}{2} \right) - I_D \left( I_{QCL} -\frac{\delta I_{QCL}}{2} \right) \right] / \delta I_{QCL}
\label{}
\end{equation} where we used \diQCL$=10$ mA. This is compared in Fig.\ \ref{fig:Fig3}(b) to the modulated diode response $\delta I_D^{AC} (I_{QCL}) / \delta I_{QCL}$ with $\omega_{mod}=1.6505$ kHz and \diQCL$ =  3.25$ mA.    With diode biases of $V_D= +0.65$ V, $V_D= +0.70$ V and $V_D= +0.75$ V both response characterization methods yield the same qualitative dependence on QCL current including a sharp increase in signal at lasing threshold followed by a rolloff as QCL current increases. Except near lasing threshold, the modulated response, $\delta I_D^{AC} (I_{QCL}) / \delta I_{QCL}$, is lower  than the quasi-static response, $\delta I_D^{AC} (I_{QCL}) / \delta I_{QCL}$ by approximately a factor of two.   Because the diode was DC voltage biased when the modulated voltage signal was measured, the lock-in signal was probably suppressed by the pinning of the diode voltage at the DC bias point.  Unfortunately, before we attempted to current bias instead of voltage bias the diode, the device failed. Characterization of a second THz transceiver under similar bias conditions indicates that the lock-in signal is 2-3 times lower under voltage bias in comparison to current bias, which may explain the discrepancy observed between the quasi-static and modulated measurements.  We also observe that the baseline of the signal below lasing threshold ($I_{QCL}<425$ mA) in both measurements grows with increasing diode bias and hypothesize that this represents the remaining thermal contribution to the diode signal because $\partial I_D / \partial T$ scales commensurately. This suggests that the modulation rate used was slower than the thermal time constant of the laser, allowing the laser to heat and cool during one period of the modulation. 

 \begin{figure}[htbp]
 \centering
   \includegraphics[keepaspectratio=true,width=2.6 in]{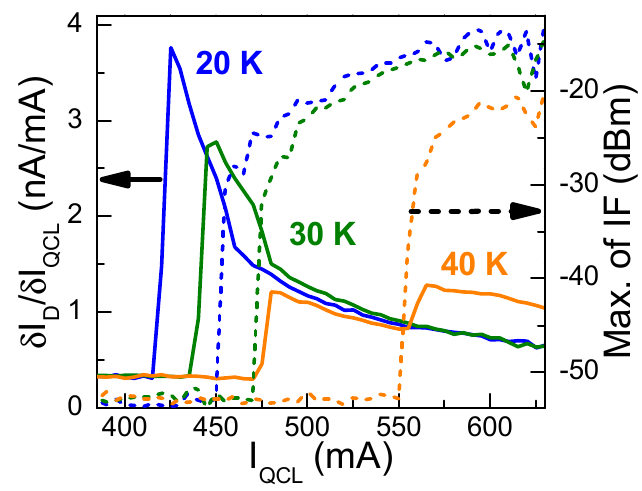}
\caption{The AC rectified diode response (solid lines) and peak of the IF signal (dashed lines) as a function of QCL current at 20 K (blue), 30 K (green) and 40 K (orange) with diode bias of 0.75 V.}
\label{fig:Fig4}
\end{figure}

Sudden jumps in the response $\delta I_D^{AC} (I_{QCL}) / \delta I_{QCL}$with AC modulation around the QCL DC bias point measured at several temperatures suggests a non-identical coupling of the fields corresponding to the different Fabry-Perot laser cavity modes to the diode.  Both the AC rectified response and amplitude of the maximum of the IF signal at 20 K, 30 K and 40 K are plotted in Fig.\ \ref{fig:Fig4}. Since the IF signal represents the mixing of two or more QCL modes, it has a higher current threshold than the DC rectified response which only requires a single lasing mode. The QCL current was modulated at 1.6505 kHz with a RMS amplitude of 3.25 mA.  There is a sudden jump in $\delta I_D^{AC} (I_{QCL}) / \delta I_{QCL}$  at the threshold current for lasing (425 mA, 450 mA and 480 mA for 20 K, 30 K and 40 K, respectively) consistent with the measurements in Fig.\ \ref{fig:Fig2}(b).  An additional set of critical features in the diode response $\delta I_D^{AC} (I_{QCL}) / \delta I_{QCL}$  is evident at IF threshold, particularly in the 40 K data where a striking increase in the rectified response is observed at the onset of multi-mode lasing.  At 20 K and 30 K there is also a noticeable but smaller decrease in the rectified response at IF threshold.  The increase in rectified response at 40 K suggests that unlike at lower temperatures, IF threshold corresponds to the onset of laser cavity modes that are more strongly coupled to the diode than the threshold mode. 

Based on Fig.\ \ref{fig:Fig4}, it is evident the diode response is not strictly proportional to the laser emission power.  The slope of the QCL power versus bias current (as shown in Fig.\ \ref{fig:Fig1}(b)) is relatively constant even after the onset of multi-mode lasing.  Yet $\delta I_D^{AC} / \delta I_{QCL}$ decreases monotonically except at IF threshold where additional structure in the signal is present.  This suggests that the internal laser mode structure has a strong effect on coupling.  Because the diode is significantly smaller than the wavelength of the F-P modes, we conclude that the strength of the coupling between the diode and laser depends upon the spatial distribution of the internal laser field. Future measurements will attempt to explore this further.

\section{Conclusion}
In this letter we have demonstrated that the response of a monolithically integrated diode embedded in a THz QCL exhibits the rectified behavior expected assuming the diode response follows the instantaneous electric field of the laser mode.  The rectified response is evident under both purely DC QCL bias conditions and with AC modulation about a DC laser bias point.  The correlation between the measured diode response and diode transport characteristics establishes that the internal laser fields couple directly to the embedded diode.

It is useful to consider the advantages of embedding a planar Schottky diode into the QCL waveguide given that an IF signal was also observed in QCLs without an integrated diode.\cite{Gellie:10}  In QCLs without an integrated diode, it remains uncertain whether the IF response results from the non-linearity of the QCL core or is due to a non-Ohmic laser contact.  Regardless of the mechanism, this non-linear response is dependent entirely upon the QCL's intrinsic characteristics.  An integrated diode can be optimized independently of the QCL, and also provides an isolated readout channel that separates the rectified and IF signals from the DC current supplied to the QCL.  The difficulty of fabrication and the perturbation of the QCL waveguide by the embedded diode are the most significant disadvantages of the approach we have studied in this article.

Contrary to expectations based upon state-of-the-art THz Schottky diode mixers,\cite{Siegel:1999je} the large parasitic capacitance present in the diode-laser circuit does not prevent coupling of laser modes with absolute frequencies well above the RC-limited bandwidth.  However, no more than several hundred $\mu$W of power exists in any single F-P mode.  High frequency Schottky diode mixers generally require several mW of LO power to optimize RF-to-IF conversion gain.\cite{Mehdi:1998kh}   While parasitics prevent ideal LO power coupling to Schottky diode mixers, it is also unlikely that the laser-diode coupling is optimized in the present generation of transceivers.  Improvement of the THz QCL transceiver performance could require higher laser power (preferably in a single mode), improved LO power coupling to the diode, or both.  To that end, further study of the coupling of the F-P modes to the embedded diode is required.  A deeper understanding of the relationship between the internal laser field distribution and laser-diode coupling strength would open additional avenues for the development of optimized THz photonic ICs with embedded mixers that have potentially wide-ranging applications in the laboratory and field.

\section*{Acknowledgments}
This work was supported by the Sandia laboratory directed research and development (LDRD) program.  Sandia National Laboratories is a multiprogram laboratory managed and operated by Sandia Corporation, a wholly owned subsidiary of Lockheed Martin Corporation,for the US Department of Energy's National Nuclear Security Administration  under contract DE-AC04-94AL85000.
 
\end{document}